\documentclass{article}

\usepackage[dblblindworkshop, final]{neurips_2026}
\workshoptitle{The 6th Workshop on Mathematical Reasoning and AI}

\usepackage[utf8]{inputenc}
\usepackage[T1]{fontenc}
\usepackage{hyperref}
\hypersetup{hidelinks}
\usepackage{url}
\usepackage{booktabs}
\usepackage{amsmath,amssymb,amsfonts}
\usepackage{nicefrac}
\usepackage{microtype}
\usepackage{xcolor}
\usepackage{graphicx}
\usepackage{array}
\usepackage{enumitem}
\usepackage{xspace}

\newcommand{\vialean}{\textsc{ViaLean}\xspace}
\newcommand{\frontier}{\mathcal{F}}
\newcommand{\atlas}{\mathcal{A}}

\title{Symbolic Search Is Not Exhausted: Persistent Proof-Space Exploration in Lean4}
\author{
Ruoran Xu\\
Xi'an Jiaotong-Liverpool University\\
\texttt{ruoran.xu24@student.xjtlu.edu.cn}
}

\begin{document}
\maketitle

\begin{abstract}
Formal theorem proving increasingly combines learned semantic guidance with verified symbolic execution. The quality of the symbolic search substrate therefore determines how much useful mathematical structure can be accumulated, reused, and exposed under a finite inference budget. We introduce \vialean, a Lean4 prover that organizes symbolic reasoning as bounded exploration of a persistent proof-state graph. Its search preserves coverage across complementary reasoning modes, merges semantically equivalent goals, retains verified intermediate structure, and observes short symbolic futures before committing to a transition. On the complete miniF2F test split, the model-free configuration solves \textbf{122/244 problems (50.0\% pass@1)}. Historical and recent symbolic reference points range from Lean's earlier \texttt{tidy} search to modern \texttt{grind} and SMT-backed verification, showing that proof-space organization remains a substantial source of capability. The same persistent state also provides a natural interface for neural--symbolic agents: neural reasoning can operate over verified regions and intermediate objects while Lean continuously expands and validates the local proof space.
\end{abstract}

\section{Introduction}

Formal theorem proving has entered a regime in which specialized language models can generate complete Lean proofs, synthesize formal data, and improve through proof-assistant feedback and reinforcement learning \citep{wang2024theoremllama,xin2024deepseek,xin2025deepseek,lin2025goedel,li2024hunyuan,lin2025goedelv2,ong2026pythagoras}. These systems place increasingly strong semantic priors over formal reasoning. Their proposals still enter a symbolic environment whose search state, reusable consequences, tactic scheduling, and verification procedures determine how efficiently finite inference is converted into a completed proof.

This observation separates two complementary sources of capability. Semantic guidance supplies mathematical jumps such as a useful lemma, witness, decomposition, or representation change. Proof-space organization determines how verified local consequences are retained, which reasoning modes remain represented, when equivalent goals are merged, and how far symbolic structure is explored before the prover commits to a branch. The first source has advanced rapidly through learned provers. The second remains an algorithmic design space inside the proof assistant.

miniF2F \citep{zheng2022minif2f} provides a useful trace of this symbolic design space. Its original deterministic Lean \texttt{tidy} baseline solves 18.0\% of the test split at pass@1. Recent evaluation reports 79/244 problems (32.4\%) for model-free Lean \texttt{grind}, while SMT-backed Dafny automation verifies 95/244 translated test problems (38.9\%) with an empty proof body \citep{baksys2026minif2fdafny}. Isabelle Sledgehammer reaches 10.4\% on miniF2F-test in its proof environment \citep{boehme2010sledgehammer,jiang2022thor}. The systems use different languages, libraries, and automation stacks, yet the progression makes one point clear: stronger symbolic organization continues to expand the set of formal problems reachable before learned proof generation is invoked.

\vialean develops this direction inside Lean~4. A proof attempt is represented as a bounded graph of semantic proof states and typed transitions. Candidate generation is partitioned into complementary symbolic perspectives so a high-volume operator family cannot monopolize the frontier. Equivalent proof states are merged and their verified structure persists across later visits. Bounded future graphs expose multi-step local consequences under saved metavariable state, and local synthesis continuously materializes exact inhabitants and useful partial applications. The resulting model-free configuration solves \textbf{122/244 = 50.0\%} of miniF2F-test at pass@1.

The system supports a broader agentic view of formal reasoning. The persistent graph can be rendered as a compact verified workspace in which a model chooses regions, proposes typed intermediate objects, or requests additional expansion. Symbolic search then converts these interventions into explicit goals, transitions, and validated consequences. This shared state connects high-level semantic reasoning to dense low-level proof development.

Our contributions are threefold. First, we formulate finite-budget Lean search as persistent proof-space exploration with explicit coverage across symbolic reasoning modes. Second, we instantiate this view in a kernel-checked prover that combines semantic state merging, bounded symbolic lookahead, continuous local synthesis, and budget-aware scheduling. Third, we report a 50.0\% model-free pass@1 result on the complete miniF2F test split and describe how the same proof state can support deeper neural--symbolic co-search. The implementation is available at \url{https://github.com/RuoranXu/ViaLean}.

\begin{figure*}[t]
    \centering
    \includegraphics[width=0.99\textwidth]{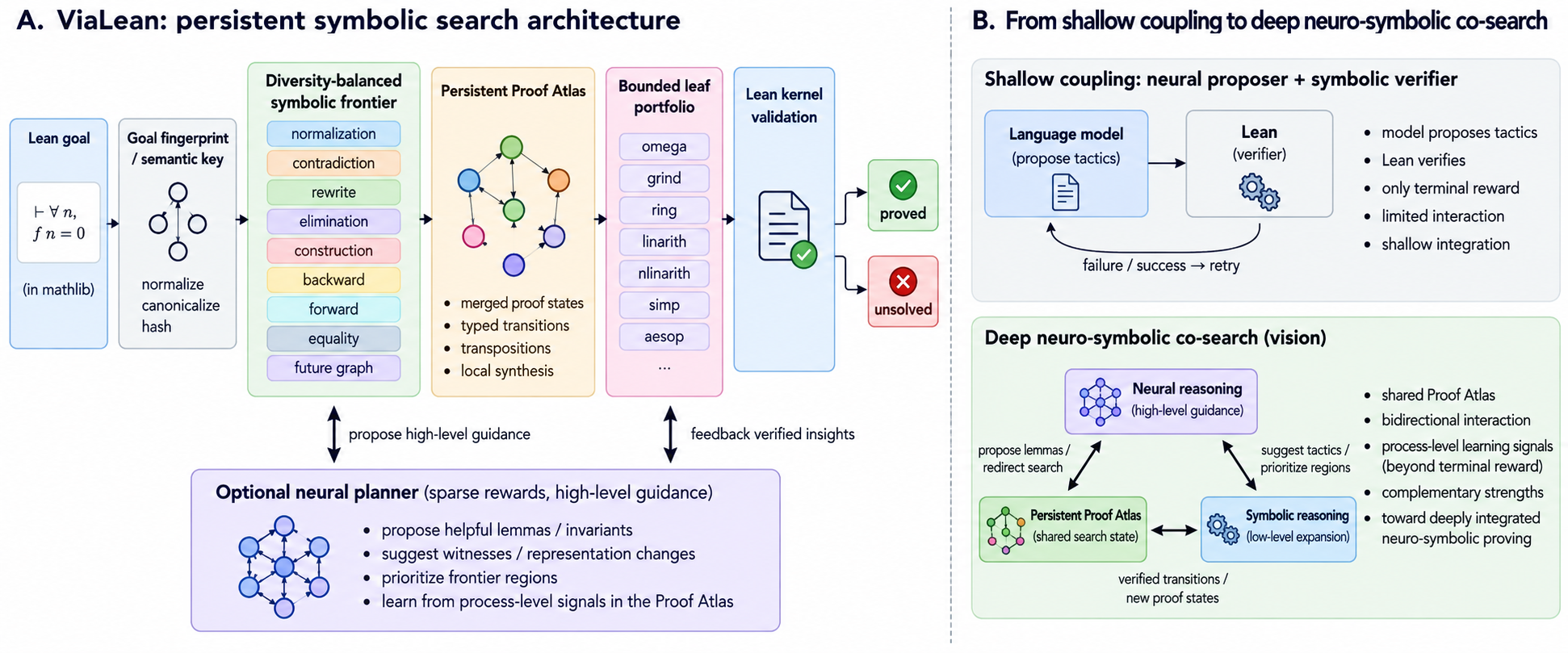}
    \caption{\textbf{Persistent proof-space exploration in \vialean.} A Lean goal is canonicalized into a semantic state, expanded through a diversity-balanced symbolic frontier, accumulated in a persistent Proof Atlas, and closed through bounded verified solvers. The Atlas also forms a shared state for optional neural guidance over regions, intermediate objects, and future symbolic expansion.}
    \label{fig:overview}
\end{figure*}

\section{Related Work and Empirical Context}

\paragraph{Symbolic proof automation.}
Interactive theorem provers have long used symbolic automation to compress routine proof development. Isabelle's Sledgehammer connects proof states to first-order and SMT solvers \citep{boehme2010sledgehammer}; Lean's mathlib provides a large formal library together with tactic-level automation \citep{mathlib2020}. Aesop develops configurable white-box best-first proof search in Lean \citep{limperg2023aesop}, and Canonical studies automated proof construction through type inhabitation and term synthesis \citep{norman2025canonical}. Across these lines, the effective search space depends on the representation of states, the scheduling of heterogeneous operators, and the amount of verified structure that survives across local attempts.

The model-free miniF2F reference points in Table~\ref{tab:baselines} capture several generations and proof environments. Lean \texttt{tidy} provides an early deterministic tactic-search baseline \citep{zheng2022minif2f}; Sledgehammer contributes mature ATP-backed automation in Isabelle \citep{jiang2022thor}; and recent measurements of Lean \texttt{grind} and Dafny illustrate the reach of modern native and SMT-backed procedures \citep{baksys2026minif2fdafny,leino2010dafny}. We use these values as contextual reference points for symbolic reach, with each row retaining its original proof environment and evaluation setting.

\paragraph{Learned Lean provers.}
Learned systems have expanded the semantic range of formal proof search. TheoremLlama specializes a general language model for Lean proof generation \citep{wang2024theoremllama}. DeepSeek-Prover and DeepSeek-Prover-V1.5 combine synthetic formal data, verifier feedback, reinforcement learning, and search \citep{xin2024deepseek,xin2025deepseek}. Goedel-Prover uses iterative data synthesis and reports 57.6\% pass@32 on miniF2F \citep{lin2025goedel}; HunyuanProver reports 68.4\% with guided tree search \citep{li2024hunyuan}; Goedel-Prover-V2 and Pythagoras-Prover further raise learned proving performance with stronger models and inference procedures \citep{lin2025goedelv2,ong2026pythagoras}. These systems motivate a symbolic substrate that exposes richer verified state than a terminal accept/reject signal, especially when models reason over several proof decisions.

\section{Persistent Proof-Space Exploration}

Let a Lean proof state be $s=(\Gamma \vdash g)$. A symbolic operation may close $g$ or produce a set of successor obligations, yielding an AND/OR transition structure. \vialean organizes this structure as a bounded Atlas $\atlas=(V,E)$ whose vertices are semantic proof states and whose edges are typed, replayable transitions. Four mechanisms determine how this Atlas grows.

\subsection{Perspective-balanced symbolic frontier}

A local proof state usually admits heterogeneous operations with very different branching factors. Rewriting and local application can generate many candidates, while constructor reasoning, contradiction, equality transport, or bounded forward closure may produce only a few. Ranking all candidates in one pool makes the candidate count of an operator family interact directly with global coverage.

\vialean partitions generation by symbolic \emph{perspective}: normalization, contradiction, rewriting, elimination, construction, backward application, typed forward closure, equality reasoning, and bounded future exploration. If $C_p(s)$ denotes candidates from perspective $p$, $q$ is a per-perspective cap, and $B$ is the global frontier budget, the merge operator is
\begin{equation}
\frontier_B(s)=\operatorname{Take}_B\!\left(
\operatorname{RR}_{p\in\mathcal P}\bigl(\operatorname{Take}_q C_p(s)\bigr)
\right),
\label{eq:frontier}
\end{equation}
where $\operatorname{RR}$ is round-robin interleaving. When $B$ covers the active perspectives, each active perspective contributes one candidate before a second candidate is drawn from any perspective. This gives a simple deterministic coverage guarantee under a fixed frontier budget and keeps multiple modes of local mathematical reasoning available to the scheduler.

\subsection{Persistent semantic states and transpositions}

Tactic histories frequently converge to the same mathematical obligation. \vialean assigns each state a strict semantic identity derived from the target and ordered local context, with local names alpha-normalized and hash matches structurally confirmed. Reaching an existing identity reuses the corresponding Atlas vertex. Its outgoing transitions, failures, synthesized objects, and verified descendants remain available to subsequent search.

This transposition structure changes the unit of accounting from a path prefix to a proof state. Repeated symbolic routes can therefore share downstream work, and verified intermediate structure becomes persistent memory within the proof attempt. The Atlas remains finite through explicit bounds on vertices, transitions, work units, meta-level operations, rendering, and wall-clock time. A single global deadline is propagated across symbolic expansion, leaf solving, recursive subgoals, and optional model interaction.

\subsection{Bounded future observation}

Many useful local transformations reveal their value after several verified steps. A rewrite may expose a normal form; an introduction may reveal a reusable local implication; a constructor can create arithmetic obligations that close immediately. \vialean explores such local futures under saved metavariable state and records a bounded future graph containing goals, operator paths, and qualitative signals. Observation is transactional: the previewed metavariable state is restored after inspection, and a chosen transition is replayed on a fresh goal before its subgoals are solved.

The future graph is controlled independently by depth, width, node, character, and work budgets. These limits make short symbolic foresight compatible with a strict global budget. The scheduler can therefore compare nearby proof regions using their verified local structure while maintaining a compact persistent workspace.

\subsection{Continuous local synthesis and verified closure}

At each visited state, a local synthesizer enumerates multiple exact inhabitants together with partial applications and their typed gaps. These objects are inserted into the same search process as tactic-generated transitions, allowing intermediate terms to reshape subsequent reachability. Shared generators also produce kernel-typed facts such as equality transport so native rewriting and model-facing state views refer to the same verified objects.

Local closure is supplied by bounded mathlib automation, including normalization, rewriting, arithmetic procedures, and general tactic search \citep{mathlib2020,limperg2023aesop}. Every successful branch is elaborated into an ordinary Lean term and checked against the requested target by Lean~4 \citep{demoura2021lean4}. The proof returned to the caller is therefore the endpoint of the same typed transition system stored in the Atlas.

\section{Model-Free Evaluation on miniF2F}

\paragraph{Protocol.}
We evaluate all 244 problems in the miniF2F test split \citep{zheng2022minif2f}. The theorem-answer declarations are excluded from the evaluation environment, and target-theorem retrieval is disabled. The evaluated configuration uses the symbolic engine, local synthesis, Atlas persistence, bounded future exploration, and mathlib leaf closure with the external model interface disabled. A case is solved when Lean constructs a complete proof term for the target. The resulting system solves \textbf{122/244 problems (50.0\% pass@1)}.

\begin{table}[t]
\centering
\caption{Selected miniF2F reference points. The symbolic rows retain their reported proof environments and evaluation settings; learned rows illustrate the higher ceiling obtained with trained prover models and larger inference budgets.}
\label{tab:baselines}
\scriptsize
\begin{tabular}{>{\raggedright\arraybackslash}p{2.45cm}>{\raggedright\arraybackslash}p{1.05cm}>{\raggedleft\arraybackslash}p{0.72cm}>{\raggedright\arraybackslash}p{2.25cm}}
\toprule
System & Regime & Test & Setting \\
\midrule
Lean \texttt{tidy} \citep{zheng2022minif2f} & symbolic & 18.0\% & pass@1, deterministic \\
Sledgehammer \citep{jiang2022thor} & symbolic & 10.4\% & Isabelle miniF2F \\
Lean \texttt{grind} \citep{baksys2026minif2fdafny} & symbolic & 32.4\% & 79/244, no model \\
Dafny verifier \citep{baksys2026minif2fdafny} & symbolic & 38.9\% & 95/244, empty proof \\
\vialean (ours) & symbolic & \textbf{50.0\%} & 122/244, pass@1 \\
\midrule
DeepSeek-V1.5-RL \citep{lin2025goedel,xin2025deepseek} & learned & 50.0\% & whole-proof pass@32 \\
Goedel-Prover-SFT \citep{lin2025goedel} & learned & 57.6\% & pass@32 \\
\bottomrule
\end{tabular}
\end{table}

\paragraph{Empirical context.}
The symbolic rows show how much verified search can be obtained from different automation stacks before learned proof generation contributes semantic proposals. ViaLean's 50.0\% result extends this sequence within Lean by combining diverse local expansion, persistent state reuse, bounded future observation, and continuous term synthesis in one search process. The learned rows occupy a different inference regime and indicate the semantic headroom available when trained priors and multiple generations are added. Table~\ref{tab:baselines} therefore separates two axes of progress: stronger proof-space organization and stronger learned semantic guidance.

The reported score is a system-level measurement of the complete model-free configuration. Appendix~\ref{app:ablations} specifies a matched-compute decomposition for isolating perspective balancing, Atlas persistence, future observation, and local synthesis under common budgets. Appendix~\ref{app:repro} records the evaluation boundary and implementation controls used to keep search, rendering, and provider work explicitly bounded.

\section{The Proof Atlas as an Agentic Interface}

A persistent proof graph creates a direct interface between symbolic search and a higher-level mathematical agent. The Atlas exposes verified states, typed transitions, synthesized objects, explored regions, and structured failures. A model can therefore condition directly on an evolving mathematical workspace whose vertices and edges summarize the verified search history.

The implementation exposes three interfaces over the same Atlas. A \emph{planner} reasons over compressed regions and can request expansion or propose typed terms, helper cuts, bridges, and witnesses; an \emph{interactive} mode receives symbolic futures and structured feedback; and a \emph{policy} mode scores existing actions. Lean elaborates proposed objects before recording their typed consequences. This gives a clean division of work: neural computation selects semantically expensive regions or objects, while symbolic search performs normalization, rewriting, elimination, construction, forward closure, synthesis, and leaf solving. Each accepted intervention updates the Atlas for the next model decision.

\subsection{Stateful neural epochs over verified search}

The interaction can be written as a sequence of stateful neural epochs. At epoch $t$, the model receives a bounded compression $o_t=\operatorname{Compress}(\atlas_t)$ containing selected Atlas regions, visible local facts, candidate transitions, future summaries, and recent structured outcomes. It returns an intervention $u_t$ drawn from a small semantic action space: prioritize a region, request expansion, propose a typed cut or bridge, supply a witness, or submit a complete candidate. Lean elaboration maps the proposal to a verified object $\widehat u_t$ when possible, after which the symbolic controller continues bounded search:
\begin{equation}
\atlas_{t+1}=\operatorname{Search}\!\left(\atlas_t,\widehat u_t;\,\mathcal B_t\right),
\label{eq:agentic_epoch}
\end{equation}
where $\mathcal B_t$ denotes the remaining node, transition, meta-operation, rendering, provider, and wall-clock budgets. The next observation is generated from the updated Atlas. Equation~\eqref{eq:agentic_epoch} makes the proof assistant's verified state the persistent carrier of information across model calls.

This interface supports long-horizon reasoning with sparse neural intervention. One model call can introduce a high-level mathematical object while the symbolic engine develops its local consequences, merges repeated states, closes tractable obligations, and returns a compact account of the updated proof region. The next neural epoch therefore receives both newly verified structure and precise information about unresolved obligations.

\subsection{Process-level supervision from Atlas deltas}

The Atlas supplies a concrete unit for process-level supervision. For an intervention $u_t$, the system can record the verified delta $\Delta\atlas_t=\atlas_{t+1}\setminus\atlas_t$ together with cost and closure events: newly reached states, proved intermediate facts, discharged obligations, transposition reuse, downstream proof depth, and symbolic work. These quantities can supervise region selection, value estimation, or proposal quality while remaining attached to kernel-checked events. Symbolic and neural improvements therefore compose through the same persistent state: stronger search develops each intervention more deeply, and stronger guidance places higher-value objects into the Atlas.

\section{Conclusion}

ViaLean organizes finite-budget theorem proving around a persistent proof-state graph with perspective-balanced expansion, semantic state reuse, bounded futures, and local synthesis. The reported model-free configuration reaches 50.0\% pass@1 on miniF2F-test. The same Atlas provides a shared verified workspace for neural--symbolic agents. Together, these mechanisms make symbolic search an active reasoning substrate with explicit state, memory, foresight, and validation. Learned configurations can use this state to carry semantic proposals and process-level feedback across model calls within one kernel-checked search trajectory.

\clearpage
\bibliographystyle{plainnat}
\bibliography{references}

@inproceedings{demoura2021lean4,
  author    = {Leonardo de Moura and Sebastian Ullrich},
  title     = {The Lean 4 Theorem Prover and Programming Language},
  booktitle = {Automated Deduction -- CADE 28},
  series    = {Lecture Notes in Computer Science},
  volume    = {12699},
  pages     = {625--635},
  publisher = {Springer},
  year      = {2021},
  doi       = {10.1007/978-3-030-79876-5_37}
}

@inproceedings{mathlib2020,
  author    = {{The mathlib Community}},
  title     = {The Lean Mathematical Library},
  booktitle = {Proceedings of the 9th ACM SIGPLAN International Conference on Certified Programs and Proofs},
  series    = {CPP 2020},
  pages     = {367--381},
  publisher = {ACM},
  year      = {2020},
  doi       = {10.1145/3372885.3373824}
}

@inproceedings{zheng2022minif2f,
  author    = {Kunhao Zheng and Jesse Michael Han and Stanislas Polu},
  title     = {MiniF2F: a Cross-System Benchmark for Formal Olympiad-Level Mathematics},
  booktitle = {International Conference on Learning Representations},
  year      = {2022}
}

@inproceedings{boehme2010sledgehammer,
  author    = {Sascha B{\"o}hme and Tobias Nipkow},
  title     = {Sledgehammer: Judgement Day},
  booktitle = {Automated Reasoning -- IJCAR 2010},
  series    = {Lecture Notes in Computer Science},
  volume    = {6173},
  pages     = {107--121},
  publisher = {Springer},
  year      = {2010},
  doi       = {10.1007/978-3-642-14203-1_9}
}

@inproceedings{jiang2022thor,
  author    = {Albert Q. Jiang and Wenda Li and Szymon Tworkowski and Konrad Czechowski and Tomasz Odrzygozdz and Piotr Milos and Yuhuai Wu and Mateja Jamnik},
  title     = {Thor: Wielding Hammers to Integrate Language Models and Automated Theorem Provers},
  booktitle = {Advances in Neural Information Processing Systems},
  volume    = {35},
  year      = {2022}
}

@inproceedings{limperg2023aesop,
  author    = {Jannis Limperg and Asta Halkj{\ae}r From},
  title     = {Aesop: White-Box Best-First Proof Search for Lean},
  booktitle = {Proceedings of the 12th ACM SIGPLAN International Conference on Certified Programs and Proofs},
  pages     = {253--266},
  year      = {2023},
  doi       = {10.1145/3573105.3575671}
}

@inproceedings{norman2025canonical,
  author    = {Chase Norman and Jeremy Avigad},
  title     = {Canonical for Automated Theorem Proving in Lean},
  booktitle = {16th International Conference on Interactive Theorem Proving (ITP 2025)},
  series    = {Leibniz International Proceedings in Informatics},
  volume    = {352},
  pages     = {14:1--14:20},
  publisher = {Schloss Dagstuhl -- Leibniz-Zentrum f{\"u}r Informatik},
  year      = {2025},
  doi       = {10.4230/LIPIcs.ITP.2025.14}
}

@inproceedings{leino2010dafny,
  author    = {K. Rustan M. Leino},
  title     = {Dafny: An Automatic Program Verifier for Functional Correctness},
  booktitle = {Logic for Programming, Artificial Intelligence, and Reasoning -- LPAR-16},
  series    = {Lecture Notes in Computer Science},
  volume    = {6355},
  pages     = {348--370},
  publisher = {Springer},
  year      = {2010},
  doi       = {10.1007/978-3-642-17511-4_20}
}

@article{baksys2026minif2fdafny,
  author  = {Mantas Baksys and Stefan Zetzsche and Olivier Bouissou and Sean B. Holden},
  title   = {MiniF2F-Dafny: LLM-Guided Mathematical Theorem Proving via Auto-Active Verification},
  journal = {arXiv preprint arXiv:2512.10187},
  year    = {2026}
}

@inproceedings{wang2024theoremllama,
  author    = {Ruida Wang and Jipeng Zhang and Yizhen Jia and Rui Pan and Shizhe Diao and Renjie Pi and Tong Zhang},
  title     = {TheoremLlama: Transforming General-Purpose LLMs into Lean4 Experts},
  booktitle = {Proceedings of the 2024 Conference on Empirical Methods in Natural Language Processing},
  pages     = {11953--11974},
  year      = {2024},
  doi       = {10.18653/v1/2024.emnlp-main.667}
}

@article{xin2024deepseek,
  author  = {Huajian Xin and Daya Guo and Zhihong Shao and Zhizhou Ren and Qihao Zhu and Bo Liu and Chong Ruan and Wenda Li and Xiaodan Liang},
  title   = {DeepSeek-Prover: Advancing Theorem Proving in LLMs through Large-Scale Synthetic Data},
  journal = {arXiv preprint arXiv:2405.14333},
  year    = {2024}
}

@inproceedings{xin2025deepseek,
  author    = {Huajian Xin and Z. Z. Ren and Junxiao Song and Zhihong Shao and Wanjia Zhao and Haocheng Wang and Bo Liu and Liyue Zhang and Xuan Lu and Qiushi Du and Wenjun Gao and Haowei Zhang and Qihao Zhu and Dejian Yang and Zhibin Gou and Z. F. Wu and Fuli Luo and Chong Ruan},
  title     = {DeepSeek-Prover-V1.5: Harnessing Proof Assistant Feedback for Reinforcement Learning and Monte-Carlo Tree Search},
  booktitle = {International Conference on Learning Representations},
  year      = {2025}
}

@article{lin2025goedel,
  author  = {Yong Lin and Shange Tang and Bohan Lyu and Jiayun Wu and Hongzhou Lin and Kaiyu Yang and Jia Li and Mengzhou Xia and Danqi Chen and Sanjeev Arora and Chi Jin},
  title   = {Goedel-Prover: A Frontier Model for Open-Source Automated Theorem Proving},
  journal = {arXiv preprint arXiv:2502.07640},
  year    = {2025}
}

@article{li2024hunyuan,
  author  = {Yang Li and Dong Du and Linfeng Song and Chen Li and Weikang Wang and Tao Yang and Haitao Mi},
  title   = {HunyuanProver: A Scalable Data Synthesis Framework and Guided Tree Search for Automated Theorem Proving},
  journal = {arXiv preprint arXiv:2412.20735},
  year    = {2024}
}

@article{lin2025goedelv2,
  author  = {Yong Lin and Shange Tang and Bohan Lyu and Ziran Yang and Jui-Hui Chung and Haoyu Zhao and Lai Jiang and Yihan Geng and Jiawei Ge and Jingruo Sun and Jiayun Wu and Jiri Gesi and Ximing Lu and David Acuna and Kaiyu Yang and Hongzhou Lin and Yejin Choi and Danqi Chen and Sanjeev Arora and Chi Jin},
  title   = {Goedel-Prover-V2: Scaling Formal Theorem Proving with Scaffolded Data Synthesis and Self-Correction},
  journal = {arXiv preprint arXiv:2508.03613},
  year    = {2025}
}

@article{ong2026pythagoras,
  author  = {Joshua Ong Jun Leang and Zheng Zhao and Mihaela Catalina Stoian and Qiyuan Xu and Haonan Li and Wenda Li and Shay B. Cohen and Eleonora Giunchiglia},
  title   = {Pythagoras-Prover: Advancing Efficient Formal Proving via Augmented Lean Formalisation},
  journal = {arXiv preprint arXiv:2606.12594},
  year    = {2026}
}
\clearpage
\appendix

\section{Semantic Proof-State Identity and Transpositions}
\label{app:goalkey}

The persistent search graph relies on a proof-state identity that is stricter than a rendered goal string.  Let a Lean state be
\[
 s=(\Gamma\vdash g),\qquad
 \Gamma=(d_1,\ldots,d_n),
\]
where each declaration $d_i$ is either a local assumption or a local let-binding.  The implementation constructs a semantic key
\begin{equation}
K(s)=\Big(\phi_{\Gamma}(g),
   \big[(\kappa_i,\phi_{<i}(\tau_i),\phi_{<i}(v_i))\big]_{i=1}^{n}\Big),
\label{eq:goal-key}
\end{equation}
where $\kappa_i\in\{\texttt{local},\texttt{letDecl}\}$ records the declaration kind, $\tau_i$ is its type, $v_i$ is its optional let-value, and $\phi$ replaces free-variable identities by position-stable bound variables and erases binder names.  The order of the local context is retained.  Implementation-detail declarations are omitted, while let-values remain semantic content.

This construction has two practical consequences.  First, alpha-renaming of user-facing binders does not create a new search state.  Second, states that agree only textually are not merged when their local declaration structure, declaration kind, type, or let-value differs.  The resulting key is used throughout cycle detection, bounded-future exploration, and Atlas transposition detection.

\paragraph{Collision-safe lookup.}
The key has a 64-bit hash bucket for fast indexing, but hash equality is not treated as semantic equality.  Membership first selects a bucket and then checks full structural equality of the key.  Thus the hash is an acceleration structure rather than the proof-state identity itself.  If $h(K_1)=h(K_2)$ while $K_1\neq K_2$, the states remain distinct.

\paragraph{Transposition insertion.}
When a newly observed goal has a key already represented in the Atlas, the existing node is reused and the transposition counter is incremented.  Otherwise a new node is allocated if the node budget remains available.  Consequently the graph can represent several proof paths entering the same semantic goal without duplicating the state itself.  This is especially useful after rewriting, constructor decomposition, or helper-lemma introduction, where different local histories can converge to the same residual obligation.

\section{Proof Atlas Data Model}
\label{app:atlas-model}

The Atlas is a typed graph with explicit evidence and bounded metadata.  A node stores
\begin{equation}
 v=(\mathrm{id},K(s),s_{\mathrm{snap}},d,p,\sigma,c),
\end{equation}
where $d$ is search depth, $p$ indicates whether the node was created by preview, $\sigma$ contains symbolic signals, and $c$ is a local difficulty estimate.  Preview nodes additionally store a durable rendered goal because the Lean metavariable state used to create them is later rolled back.  A transition stores
\begin{equation}
 e=(\mathrm{id},v_{\mathrm{src}},a,\{v_j\},x,z,\epsilon),
\end{equation}
with symbolic candidate $a$, child nodes $\{v_j\}$, executability flag $x$, coupling flag $z$, and transition evidence $\epsilon$.  Evidence records one of \texttt{untried}, \texttt{expanded}, \texttt{solved}, \texttt{failed}, \texttt{timedOut}, or \texttt{rejected}, together with elapsed time and an optional failure class.

The symbolic candidate itself is separated into three layers:
\begin{enumerate}[leftmargin=1.7em,itemsep=2pt,topsep=3pt]
    \item a concrete operation, such as local application, constructor, cases, induction, rewrite, bridge, witness, cut, leaf solver, or exact term;
    \item a coarse strategy family used by scheduling and region aggregation;
    \item metadata consisting of prior, estimated cost, origin, and a stable fingerprint.
\end{enumerate}
The strategy families are normalization, contradiction, equality, elimination, construction, backward, forward, witness, cut, structural, leaf, and mixed.  The same vocabulary is shared by symbolic producers, the Atlas, the scheduler, and the planner interface.

\subsection{Transition uniqueness and evidence updates}

For a fixed parent node, a transition fingerprint acts as the identity of a candidate edge.  Offering an already present fingerprint returns the existing transition.  A new edge is inserted only while both the transition-count budget and the work-unit budget remain open.  Successful child attachment marks the transition as expanded; a solved edge records a terminal positive outcome; failed or rejected outcomes mark its knowledge status as refuted.  These status updates affect search memory and planner observations, not Lean's proof-theoretic acceptance criterion.

\subsection{Structural transition potential}

For previews with children, the implementation computes a bounded structural value from parent and child difficulty estimates.  Let $c_p$ be the parent cost, $c_1,\ldots,c_m$ the child costs, and
\[
 \bar c=\frac{1}{m}\sum_{j=1}^{m}c_j,
 \qquad b(m)=0.20(m-1).
\]
Then an expanded transition receives
\begin{equation}
 V_{\mathrm{sym}}(e)=
 \operatorname{clip}_{[0,1]}
 \frac{c_p}{\max\{0.1,c_p+\bar c+b(m)\}},
\label{eq:transition-potential}
\end{equation}
while a solved edge has value $1$ and an unexpanded edge has no structural value.  The branch penalty prevents a decomposition into many individually small children from appearing automatically preferable.  This quantity is a search signal; it is never used as evidence that a proposition is true.

\subsection{Coarse proof regions}

The Atlas can be compressed into proof regions indexed by strategy family.  For each family, the region stores its entry nodes, a small set of depth-ordered representatives, size, depth range, and aggregate signals such as solved edges, failed edges, and nodes admitting an exact local inhabitant.  Region compression provides a stable intermediate granularity between a single tactic action and the entire proof graph.  Planner mode operates primarily at this granularity, while the underlying nodes and transitions remain available when additional detail is needed.

\section{Perspective-Balanced Frontier Construction}
\label{app:frontier-detail}

At a state $s$, each symbolic perspective $p\in\mathcal P$ independently emits an ordered sequence
\[
 C_p(s)=(a_{p,1},a_{p,2},\ldots).
\]
The current implementation covers normalization, contradiction, equality/rewrite, elimination, construction, backward application, forward reasoning, and multi-step symbolic futures, together with other proposal families from the main search controller.  Let $q$ denote the per-perspective cap and $B$ the visible frontier budget.  Define
\[
 \widetilde C_p(s)=C_p(s)_{1:\min(q,|C_p(s)|)}.
\]
The global frontier is formed by round-robin interleaving of the capped sequences and truncation to $B$ items, as stated in Eq.~\eqref{eq:frontier}.

\paragraph{First-round coverage property.}
Let $A(s)=\{p:C_p(s)\neq\varnothing\}$ be the active perspectives.  If $q\ge 1$ and $B\ge |A(s)|$, the merged frontier contains at least one candidate from every active perspective before any perspective contributes its second candidate.  This follows because the first interleaving round visits each active sequence once.  Candidate multiplicity inside one family therefore cannot remove all first-round representation from a lower-cardinality family.

\paragraph{Why the cap is separate from the global budget.}
The per-perspective cap controls domination by a prolific producer; the global budget controls the visible action set.  Keeping them separate allows a family to remain represented without forcing the total action set to scale with the raw number of rewrites, local applications, or retrieved premises.

\section{Bounded Symbolic Future Graph}
\label{app:future-graph}

The future graph is an observation mechanism that expands short symbolic trajectories under saved Lean metavariable state.  The implementation uses breadth-first exploration.  Each queue item contains a goal metavariable, a depth, and an operator path.  A node contributes a rendered goal and qualitative symbolic signals; if it remains within the depth and node budgets, the engine tries at most one successful successor group from each eligible operator family before further descendants are expanded.

A compact description is:
\begin{enumerate}[leftmargin=1.7em,itemsep=2pt,topsep=3pt]
    \item initialize the queue with the current goal and insert its semantic key into a strict visited set;
    \item pop queue items in FIFO order until the node cap, work cap, or wall-clock budget closes;
    \item render the current state and attach local signals;
    \item generate bounded successor groups from introduction, rewriting, elimination, induction at the root, construction, and backward application;
    \item for each accepted successor group, create a typed preview transition and bounded child states;
    \item deduplicate children using the same strict goal identity and enqueue unseen children while depth remains;
    \item restore Lean's metavariable state after the observation transaction, while retaining the resulting durable Atlas records.
\end{enumerate}

The breadth-first policy is important because the future view is itself a scarce observation budget.  A depth-first procedure can spend most of the budget following one operator family.  FIFO expansion instead exposes competing local futures at comparable depths.  Within each future node, successor groups are sorted by estimated symbolic cost with deterministic fingerprint tie-breaking; a requested strategy family can be moved to the front without discarding the alternatives.

\subsection{Preview versus execution}

Future expansion and proof execution are intentionally distinct.  Preview operations run in a saved metavariable state and can create durable descriptions of counterfactual states, but they do not commit a root proof assignment.  When the controller selects an executable transition, the operation is replayed on a fresh goal and its recursive obligations are solved through the ordinary controller.  This separation permits richer observation without making speculative state mutation part of the trusted proof path.

\subsection{Finite observation bounds}

The future view has independent caps for depth, per-node width, total future nodes, branch fan-out, rendered characters, and meta-level work.  With defaults from the public configuration, these are depth $3$, width $6$, at most $24$ future nodes, and at most $6$ displayed/replayed children per branch.  These are implementation defaults rather than claims about the benchmark configuration; the harness may override them.

\section{Continuous Local Synthesis}
\label{app:synthesis}

Local synthesis runs at visited proof states and exposes complete and partial inhabitants independently of model guidance.  For each local declaration $h:\tau$, the synthesizer first checks whether $\tau$ is definitionally equal to the target $g$.  If so, it emits a zero-cost exact inhabitant.  Otherwise it follows a bounded chain of nondependent arrows in $\tau$ and records the resulting domains as obligations.  The candidate is represented as
\[
 q=(t,\Omega,\mathrm{op},c),
\]
where $t$ is the local term, $\Omega$ is the array of missing obligations, $\mathrm{op}$ is the corresponding symbolic operation, and $c=|\Omega|$ for partial local applications.  A candidate is complete exactly when $\Omega=\varnothing$.

The current synthesizer also detects reflexive equality goals and the proposition \texttt{True}.  For an equality $x=x$ up to definitional equality it materializes \texttt{Eq.refl}; for \texttt{True} it materializes \texttt{True.intro}.  These small rules matter because complete inhabitants become direct closure opportunities, while partial inhabitants expose useful typed decompositions to later search.

The synthesis layer is deliberately typed.  It does not serialize arbitrary source-level terms as proof facts.  The term and its obligations remain Lean expressions inside the proof process, and any complete candidate still passes the final validation path before the root proof is returned.

\section{Transactional Execution and Meta-State Isolation}
\label{app:transactions}

Lean metaprogramming manipulates mutable metavariable assignments, so speculative search requires explicit isolation.  ViaLean uses a transactional pattern at the operation boundary:
\begin{enumerate}[leftmargin=1.7em,itemsep=2pt,topsep=3pt]
    \item save the complete metavariable state before preview or branch-local experimentation;
    \item execute a bounded transformation under the current local context;
    \item extract only stable information needed by the Atlas, such as rendered child goals, typed transition metadata, and structural signals;
    \item restore the saved state after preview, failed execution, or benchmark isolation;
    \item commit the original root assignment only after a complete candidate has been recursively solved and validated.
\end{enumerate}

This creates two memory layers.  The \emph{ephemeral Meta state} contains temporary metavariables and trial assignments.  The \emph{persistent workspace} contains semantic keys, durable node descriptions, transition evidence, verified objects, structured failures, and version numbers.  Rollback therefore removes speculative Lean assignments without discarding what the search learned about the local proof space.

\paragraph{Nested model calls.}
Model-generated subgoals and selected replay operations execute with nested model queries disabled.  This keeps one semantic intervention from recursively spawning uncontrolled provider interactions and preserves a single accounting path for the outer search episode.

\section{Budget Algebra and Operational Boundedness}
\label{app:budgets}

The search budget is a vector rather than one scalar:
\begin{equation}
 \mathcal B=(T,N,E,W,M,C,R,D,F,P),
\end{equation}
where $T$ is wall-clock time, $N$ Atlas nodes, $E$ Atlas transitions, $W$ work units, $M$ Meta operations, $C$ rendered characters, $R$ regions, $D$ future depth, $F$ future nodes/width, and $P$ provider/planner capacity.  Different producers debit different coordinates.  In particular, increasing the number of characters shown to a model does not create new symbolic work units, and increasing a displayed frontier cap does not silently increase the persistent Meta-operation allowance.

\begin{table}[h]
\centering
\small
\begin{tabular}{@{}p{0.33\linewidth}r p{0.46\linewidth}@{}}
\toprule
Configuration field & Default & Role \\
\midrule
\texttt{timeoutSec} & 10 & shared wall-clock deadline \\
\texttt{maxDepth} & 2 & controller recursion depth \\
\texttt{maxActionsPerNode} & 32 & executable action cap \\
\texttt{frontierMaxProbes} & 32 & visible frontier probes \\
\texttt{frontierMaxPerPerspective} & 4 & per-perspective visibility cap \\
\texttt{frontierFutureDepth} & 3 & symbolic future depth \\
\texttt{frontierFutureWidth} & 6 & successor groups per future node \\
\texttt{frontierFutureNodes} & 24 & total future-view nodes \\
\texttt{atlasMaxNodes} & 96 & persistent semantic states \\
\texttt{atlasMaxTransitions} & 160 & persistent typed transitions \\
\texttt{atlasMaxWorkUnits} & 256 & Atlas transition work cap \\
\texttt{atlasMaxMetaOps} & 256 & shared Meta-operation cap \\
\texttt{atlasMaxRenderedChars} & 16000 & persistent rendering budget \\
\texttt{atlasMaxRegions} & 12 & compressed planner regions \\
\texttt{plannerMaxCalls} & 4 & planner calls in one search \\
\texttt{plannerMaxPayloadChars} & 16000 & serialized planner payload cap \\
\bottomrule
\end{tabular}
\caption{Selected public defaults from the implementation.  Benchmark or application configurations may override these values.}
\label{tab:default-budgets}
\end{table}

The global deadline is installed as a Lean cancellation token over the complete search.  The monitor observes the remaining time and cancels provider, Atlas, retrieval, synthesis, and solver computation through the same outer boundary.  Local timeout cancellation is converted into an ordinary failed attempt, while an already-set parent cancellation is propagated.

\paragraph{Operational boundedness.}
At the controller level, all explicit queues and loops have finite count or work caps, and the full operation is covered by the shared cancellation deadline.  Thus the search cannot expand its own Atlas, future queue, provider loop, or trace stream without bound.  This statement concerns the implementation's search control; it is distinct from a complexity theorem about arbitrary Lean elaboration.

\section{Planner and Interactive Protocols}
\label{app:protocol}

The persistent workspace supports three model-facing modes over the same search state.

\paragraph{Policy mode.}
A model receives a set of kernel-checkable actions and returns bounded scalar signals.  These signals modify ranking only.  They cannot close a goal or create a verified proposition.

\paragraph{Interactive mode.}
The request contains the current goal shape, local context summaries, actions, symbolic frontier probes, bounded future views, and structured feedback from previous non-model search events.  A continuation may select actions or probes and may include Lean candidates.  Feedback records action family, outcome, elapsed time, and a bounded detail string.

\paragraph{Planner mode.}
The planner receives a compressed workspace containing a root view, proof regions, selected nodes, typed transitions, recent observations, verified/speculative objects, remaining wall-clock time, and remaining Atlas work.  Its response can include region preferences, transition policy/value/confidence values, a multi-family strategy, expansion requests, typed thought objects, and Lean candidates.

One planner epoch can be written as
\begin{equation}
 (\pi_t,\rho_t,U_t)=\operatorname{Plan}(\operatorname{Compress}(\atlas_t),\mathcal B_t),
\end{equation}
where $\pi_t$ contains transition-level signals, $\rho_t$ contains region-level or expansion decisions, and $U_t$ contains proposed intermediate objects.  The controller validates $U_t$, performs bounded symbolic work, and produces an updated workspace version.  The next request includes the corresponding Atlas delta and observations.

\subsection{Deterministic payload degradation}

Both interactive and planner requests have absolute character caps.  When a request exceeds the cap, the serializer applies a deterministic degradation schedule rather than relying on provider truncation.  Planner requests first remove older observations, then reduce object, node, transition, and region collections, and finally shorten the rendered root goal.  If the minimal request still does not fit, the payload becomes an empty JSON object.  This keeps provider work explicitly bounded and makes truncation behavior reproducible.

\section{Typed Proposal Validation}
\label{app:validation}

Model outputs and internally generated bridges cross a narrow validation boundary before they can become verified workspace objects.  The common validator instantiates metavariables and rejects an expression if it still contains unresolved metavariables, contains a loose bound variable, exceeds the configured proposal-size cap, or fails type inference.  Specialized validators then enforce the target-dependent typing relation.

\begin{table}[h]
\centering
\small
\begin{tabular}{@{}p{0.23\linewidth}p{0.69\linewidth}@{}}
\toprule
Proposal type & Validation condition \\
\midrule
Helper cut & candidate is a well-typed sort; by default it must be a proposition; it differs definitionally from the target, \texttt{True}, and existing local assumptions \\
Equality midpoint & target has form $x=y$ over carrier $\tau$; candidate has type $\tau$ and differs definitionally from both endpoints \\
Iff midpoint & target has form $P\leftrightarrow Q$; candidate is a proposition and differs definitionally from $P$ and $Q$ \\
Existential witness & target has form $\exists x:\tau, P(x)$; candidate has type $\tau$ \\
Raw Lean candidate & accepted only through the explicitly enabled experimental path, with independent candidate count, source-size, syntax, and heartbeat limits \\
\bottomrule
\end{tabular}
\caption{Target-specific proposal checks before an object can influence verified search.}
\label{tab:validation}
\end{table}

Validation does not by itself prove a helper proposition.  A well-typed cut remains an obligation until its proof is constructed.  Likewise, a witness merely instantiates an existential and leaves the instantiated proposition as a proof obligation.  The trust boundary is therefore defined by Lean elaboration and final kernel checking, not by model confidence or proposal syntax.

\section{Knowledge Objects and Workspace Deltas}
\label{app:objects}

In addition to proof states, the workspace stores typed intermediate objects.  An object records an identifier, conjecture kind, expression, optional type, blockers, utility, dependencies, knowledge status, origin, and display text.  Supported kinds include helper lemma, equality bridge, iff bridge, witness, invariant, generalization, and term.  Status is one of verified, pending, speculative, refuted, or dominated.

A planner thought batch enters the workspace as speculative objects.  Each thought is absorbed independently so one malformed or later-refuted object does not invalidate unrelated proposals.  After elaboration or proof construction, the object's status can be upgraded to verified or changed to refuted.  Duplicate objects are merged by kind, expression, and optional type, preserving the stronger knowledge status.

The workspace is versioned.  A delta between versions can list newly added nodes, transitions, verified/refuted objects, and structured observations.  This versioning provides a natural event boundary for sparse neural intervention: the planner does not need the full historical token sequence to know what changed since its previous call.

\section{Failure Taxonomy and Structured Feedback}
\label{app:failures}

Failed search is represented explicitly.  The current failure classes include type mismatch, unification failure, no progress, branch growth, branch explosion, timeout, unsafe syntax, rewrite mismatch, premise mismatch, open goals, invalid planner selection, budget denial, internal failure, and an unknown fallback.  A structured observation links the failure class to an optional transition, transition outcome, newly reached nodes, elapsed time, and a bounded debug field.

This distinction matters for both scheduling and training data.  A timeout indicates that an operation exceeded a resource boundary; a type mismatch indicates semantic incompatibility; branch explosion indicates unfavorable search geometry; and budget denial means the proposal was never executed.  Collapsing all four into a single negative reward would erase information that can guide later region selection or curriculum construction.

\section{Matched-Compute Benchmark Instrumentation}
\label{app:benchmark-modes}
\label{app:ablations}

The repository contains a matched-compute benchmark interface with eight modes.  All modes inherit the same wall-clock and Atlas work caps from a common base configuration, while the target feature is toggled:
\begin{enumerate}[leftmargin=1.7em,itemsep=2pt,topsep=3pt]
    \item \texttt{native-only}: bounded native solver without the symbolic frontier or Atlas graph;
    \item \texttt{symbolic-actions}: symbolic actions without frontier or graph persistence;
    \item \texttt{flat-frontier}: frontier exposure with deterministic replay and no Atlas graph;
    \item \texttt{graph-atlas}: model-free frontier plus persistent graph;
    \item \texttt{atlas-policy}: graph plus deterministic policy-style guidance;
    \item \texttt{atlas-policy-value}: planner-style policy/value signal without expansion requests;
    \item \texttt{planner-expansion}: planner-guided region expansion under the same outer budget;
    \item \texttt{interactive-raw}: interactive path with the explicitly enabled raw Lean compatibility gate.
\end{enumerate}

Each case emits a \texttt{vialean.benchmark.v3} record containing solved status, elapsed milliseconds, direct and proposal attempts, bridge depth, model calls, replan count, Atlas nodes, Atlas transitions, and Atlas Meta operations.  These measurements make it possible to separate accuracy gains from simple increases in symbolic work or provider calls.

For the paper-level system analysis, the most informative matched-compute contrasts are flat frontier versus graph Atlas, and graph Atlas versus full future/synthesis search.  Useful secondary statistics are transposition count, unique semantic states per solved case, successful transition depth, and time-to-proof quantiles.  No additional empirical numbers are inserted here; the reported miniF2F result in the main text remains the manuscript's measured result.

\section{Training and Analysis Traces}
\label{app:traces}

The training trace schema is intentionally different from a raw proof transcript.  A \texttt{vialean.training.v3} event contains a sequence number, event kind, workspace version, and a JSON payload.  The writer is deterministic and caps the number of serialized events.  Callers are expected to record stable identifiers, counts, and scores; raw goals, local names, provider credentials, and source-level proof text are excluded from this trace format.

This event design supports process analysis without making the trace itself part of the trusted proof artifact.  For example, a training pipeline can correlate a planner intervention with later node growth, solved transitions, bridge depth, or failure classes while the final proof remains an ordinary Lean term checked independently of the trace.

\section{Reproducibility and Evaluation Boundary}
\label{app:repro}

The manuscript's miniF2F evaluation covers the complete 244-problem test split used in the original experiment.  The theorem-answer declarations are excluded and retrieval of the target theorem is disabled.  Each successful case must terminate in a complete Lean proof term accepted by the kernel.  The reported pass@1 value is therefore $122/244=50.0\%$, exactly as in the original empirical result.

ViaLean's standalone core uses ordinary Lean~4 metaprogramming APIs.  Mathlib support is isolated in the integration project, which supplies the benchmark adapter and bounded leaf portfolio.  The search controller uses one shared deadline across frontier expansion, local synthesis, leaf solving, recursive execution, and optional provider interaction.  Proof states are keyed by the strict semantic identity in Appendix~\ref{app:goalkey}, with structural equality confirmation after hash lookup.

\paragraph{Suggested reproduction record.}
For each test theorem, a reproducible run should store the repository revision, Lean and mathlib revisions, complete configuration, timeout, solved status, elapsed time, Atlas node/transition/Meta-op counts, and the final proof term or replayable proof script.  For any model-enabled variant, the provider/model identifier, decoding parameters, request cap, model-call count, and replay data should also be retained.

\section{Baseline Comparability}
\label{app:baseline-comparability}

Table~\ref{tab:baselines} preserves the settings reported by each source.  Lean \texttt{tidy} and ViaLean are pass@1 Lean searches from different toolchain generations.  The reported \texttt{grind} number is a direct Lean tactic reference point.  Sledgehammer is evaluated in Isabelle.  Dafny uses an auto-active verifier and SMT backend over a translated miniF2F benchmark.  The learned rows use larger pass@k or search budgets.  The table therefore provides empirical context along symbolic-automation and learned-guidance axes while retaining the distinctions among proof environments and inference budgets.

\section{Implementation Map}
\label{app:implementation-map}

The technical components described above correspond directly to modules in the public implementation:
\begin{table}[h]
\centering
\small
\begin{tabular}{@{}p{0.33\linewidth}p{0.58\linewidth}@{}}
\toprule
Module & Primary responsibility \\
\midrule
\texttt{GoalKey.lean} & alpha-stable strict proof-state identity and collision-safe strict goal set \\
\texttt{Symbolic.lean} & common strategy-family and symbolic-operation intermediate representation \\
\texttt{Atlas/Types.lean} & nodes, transitions, evidence, regions, limits, transpositions, and structural potential \\
\texttt{Workspace.lean} & versioned persistent Atlas, typed objects, observations, deltas, and failure classes \\
\texttt{Frontier.lean} & perspective probes, bounded future graph, breadth-first expansion, and preview transactions \\
\texttt{Synthesis/Local.lean} & exact inhabitants and typed partial local applications \\
\texttt{Model/Protocol.lean} & policy, interactive, and planner schemas with deterministic payload caps \\
\texttt{Validate.lean} & common and target-specific validation for cuts, bridges, and witnesses \\
\texttt{Search/Controller.lean} & shared cancellation deadline over the complete search \\
\texttt{Benchmark.lean} & matched-compute modes and \texttt{vialean.benchmark.v3} records \\
\texttt{Trace.lean} & deterministic redacted \texttt{vialean.training.v3} event stream \\
\bottomrule
\end{tabular}
\caption{Implementation correspondence for the technical appendix.}
\label{tab:module-map}
\end{table}

The central design invariant across these modules is that search memory may be rich, persistent, and model-facing, while proof acceptance remains narrow: a result is returned only when Lean has a complete, well-typed term for the original target with no unresolved synthetic metavariables.

\end{document}